\documentclass[aps,a4paper,twocolumn,superscriptaddress]{revtex4}

\usepackage[american]{babel}
\usepackage[latin1]{inputenc}
\usepackage{geometry}
\usepackage{textcomp}
\usepackage{graphicx}
\usepackage{bm}
\usepackage{units}

\renewcommand{\vec}{\bm}

\newcommand{\TUKL}{Department of Physics and Research Center OPTIMAS, University of Kaiserslautern, 67663 Kaiserslautern, Germany}
\newcommand{\IPM}{Fraunhofer Institute for Physical Measurement Techniques IPM, 79110 Freiburg, Germany}

\newcommand{\epseff}{\epsilon_\mathrm{eff}}
\newcommand{\mueff}{\mu_\mathrm{eff}}
\newcommand{µ}{\textnormal{\textmu}}
\newcommand{\im}{\mathrm{i}}
\newcommand{\epso}{\epsilon_\mathrm{o}}
\newcommand{\muo}{\mu_\mathrm{o}}

\begin{document}

\title{Experimental and numerical studies of terahertz surface waves on a thin metamaterial film}

\author{Benjamin Reinhard}\email{breinhard@physik.uni-kl.de}\affiliation{\TUKL}
\author{Oliver Paul}\affiliation{\TUKL}
\author{René Beigang}\affiliation{\TUKL}\affiliation{\IPM}
\author{Marco Rahm}\affiliation{\TUKL}\affiliation{\IPM}

\begin{abstract}%
\noindent
We present experimental and numerical studies of localized terahertz surface waves on a subwavelength-thick metamaterial film consisting of in-plane split-ring resonators. A simple and intuitive model is derived that describes the propagation of surface waves as guided modes in a waveguide filled with a Lorentz-like medium. The effective medium model allows to deduce the dispersion relation of the surface waves in excellent agreement with the numerical data obtained from 3-D full-wave calculations. Both the accuracy of the analytical model and the numerical calculations are confirmed by spectroscopic terahertz time domain measurements.
\end{abstract}

\maketitle

\noindent
Metamaterials have gained a great deal of interest in the last decade due to their important role as designer materials with tailorable electric and magnetic properties. In this context, metamaterials provide a comprehensive tool box for the design of optical components with a very specific and in some cases even exotic electromagnetic behavior. From the conceptional point of view, the development of such optics requires the design of individual subwavelength-sized unit cells. When assembled properly, these subwavelength units stamp their electromagnetic characteristics on the effective macroscopic properties of the resulting metamaterial. That way, it is possible to create optics with a very specific functionality.

While free-space optics requires three-dimensional bulk metamaterials in most of the cases, it was shown that the surfaces of (thin) metamaterials (also termed meta-surfaces) can support the propagation of surface waves under well-defined conditions. Compared to the highly sophisticated methods that must be employed for the fabrication of bulk metamaterials, the techniques for the fabrication of most types of meta-surfaces are considerably less involved. As an intriguing analogy, the observed surface waves on meta-surfaces are very similar in their properties to surface plasmon polaritons on interfaces between a metal and a dielectric. In contrast to metals yet, the meta-surfaces can be specifically designed as to their effective material parameters. In consequence, it is possible to tune the dispersion characteristics and spatio-temporal behavior of surface waves on metasurfaces by design.

For example, it has already been successfully demonstrated that the structuring of metal surfaces by small subwavelength slits or holes can lead to an improved confinement of surface waves. By this method, it was possible to localize surface plasmon polaritons near the metal surfaces at terahertz (THz) and microwave frequencies where metals almost behave like perfect electric conductors \cite{pendry2004,hibbins2005,williams2008}. The same concept was applied to the design of waveguides that consisted of structured metal sheets \cite{ulrich1973,shen2005,zhu2008b}. Recently, in a similar approach, a waveguide structure based on complemetary split-ring resonators was proposed \cite{navarro-cia2009}. In an early experimental work, it was shown that magnetic surface plasmons can exist on metamaterials with negative permeability \cite{gollub2005}. Moreover, it was discussed in a theoretical work that materials with simultaneously negative permittivity and permeability can also sustain surface polaritons \cite{ruppin2000}. Facing the possibility of designing meta-surfaces with well-defined spatio-temporal and spectral properties, it becomes obvious that such metamaterials offer the possibility to create compact, integrated plasmonic devices. However, to exploit the full potential of meta-surfaces with respect to the design process of tailored plasmons, it is necessary to understand the physical mechanisms in the interaction of meta-surfaces and surface waves.

\begin{figure}%
\centering%
\includegraphics{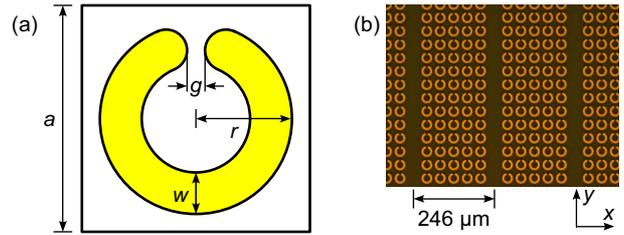}%
\caption{(a) Schematic of one unit cell of the split-ring array, $a=\unit[41]{µm}$, $r=\unit[17.25]{µm}$, $w=\unit[7.5]{µm}$, $g=\unit[3.25]{µm}$. (b) Microscope image of a sample with grating structure.}%
\label{fig:sample}%
\end{figure}

In this letter, we experimentally demonstrate the excitation of terahertz surface waves on thin magnetic metamaterial films and provide a simple, intuitive model to describe the effective properties of the meta-surface. While previous experimental investigations in the THz regime focussed on structured metal sheets \cite{ulrich1973,zhu2008b,williams2008}, the magnetic metamaterials employed in this work consisted of a single layer of in-plane split-ring resonators (SRRs). The unit cell size of the structure was $\unit[41]{µm}\times\unit[41]{µm}$, which corresponds to approx.\ $\lambda/7$ at a frequency of \unit[1]{THz}. The dimensions of the rings are shown in Fig.~\ref{fig:sample}(a). The THz surface waves were excited by magnetic field coupling between the THz radiation and the magnetic resonances of the SRRs. It should be noted that the magnetic field of the guided mode is polarized normal to the surface in strict contrast to surface plasmons on metals.

The meta-surfaces were fabricated by standard UV lithography. Details on the fabrication methods can be found in \cite{paul2008}. The copper SRRs were completely embedded in benzocyclobutene (BCB), a low-loss dielectric with a refractive index of $n=1.63$. The silicon substrate which was used during the fabrication process was removed in a post-process. The resulting samples were free-standing metamaterial membranes with a total thickness of approximately \unit[50]{µm}. An exemplary microscope image of a sample is shown in Fig.~\ref{fig:sample}(b).
Since the dispersion relation of the localized surface waves is settled to the right of the light line of the surrounding material (air), incident THz waves cannot be directly coupled to the surface modes of the metamaterial structure. Therefore, we created an artificial grating coupler by skipping every sixth row of split rings in $x$ direction to convert THz waves to bound surface states. Due to diffraction at the grating structure, the incident THz waves experience an additional momentum of $k_x = m\times 2\pi/(6a)$ ($m$: order of diffraction, $6a$: grating period) parallel to the surface of the metamaterial. This allows that both the momentum and the energy are conserved in the conversion process if the frequency of the incident THz wave is below a specific threshold value given by the grating period $6a$. In vacuum, for example, a parallel momentum $k_x = 2\pi/(6a)$ with $a=\unit[41]{µm}$ corresponds to a frequency of $\nu \approx \unit[1.22]{THz}$. This means that below this frequency, the momentum of the first-order diffracted beam parallel to the surface is above threshold and the THz wave can couple to surface modes of the meta-surface.

\begin{figure}%
\centering%
\includegraphics{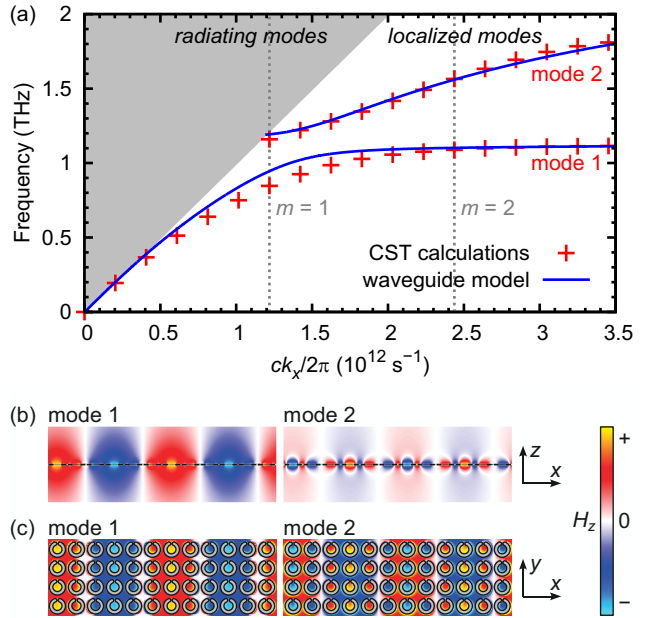}
\caption{(a) Dispersion relation of the localized wave. (b) Magnetic field normal to the metamaterial surface ($z$ component) in the $x$--$z$ plane and (c) in the $x$--$y$ plane.}%
\label{fig:dispersion}%
\end{figure}

To gain a basic understanding of the propagation of localized surface modes on magnetic meta-surfaces, we developed a simple and intuitive model. For this purpose, we treated the metasurface as a thin slab waveguide with effective electric and magnetic material parameters. In the model, the waveguide extended from $x,y=-\infty \ldots \infty$ and was confined in the $z$-direction between $-d<z<+d$ such that the slab waveguide thickness was $2d$. We assumed that the material was homogeneous and isotropic. To account for the resonant magnetic response of the split-ring resonators, we described the effective permeability $\mueff$ using a Lorentz model
\begin{equation}%
\mueff = 1 + \frac{F_1\omega^2}{\omega_1^2-\omega^2-\im\Gamma_1\omega}+ \frac{F_2\omega^2}{\omega_2^2-\omega^2-\im\Gamma_2\omega}%
\end{equation}
with resonance frequencies $\omega_j$, geometry factors $F_j$, and collision frequencies $\Gamma_j$ \cite{pendry1999}.
The effective permittivity $\epseff$ was kept at a constant value that was equal to the permittivity of BCB.
The material outside the slab was described by a permittivity $\epso$ and permeability $\muo$ ($\epso \approx \muo \approx 1$ in air).
To obtain the electromagnetic modes inside the waveguide, we considered a plane wave
\begin{equation}%
\vec E = \vec E_0 \exp{\left[\im (k_x x + k_z z - \omega t)\right]}%
\end{equation}
propagating in the $x$--$z$ plane with a wave vector $\vec k = (k_x, k_z)$ and a frequency $\omega$. It should be noted that in this case $k_x$ is a complex quantity describing the phase propagation and absorption of the wave. The characteristics of the guided modes in the waveguide are mainly determined by the reflection at the boundaries between the slab material and the surrounding material. 
A stable mode pattern can only exist if the phase advance after two reflections at opposite boundaries of the material slab is an integer multiple of $2\pi$.
Since a meta-surface only consists of a single layer of resonant elements, only mode distributions with a single maximum of the field amplitude in the center of the waveguide are meaningful solutions. These modes are obtained if the phase advance corresponds to $1\times2\pi$ per round trip. Also, we only considered transverse electric (TE) polarized modes (electric field vector points in $y$ direction) to correctly account for the magnetic response of the SRRs in $z$-direction.
For TE polarization, the Fresnel coefficient of reflection is given by
\begin{equation}%
\rho = \frac{\muo k_z - \mueff k_{\mathrm{o}z}}{\muo k_z + \mueff k_{\mathrm{o}z}}\quad.%
\end{equation}
The wave number $k_{\mathrm{o}z}$ is determined by $k_{\mathrm{o}z} = [\epso\muo(\omega/c)^2-k_x^2]^{1/2}$ and the sign of the square root is chosen such that $|\rho| \leq 1$.
Finally, the condition for the dispersion relation of the localized modes can be written as
\begin{equation}%
\arg{(\rho^2)}+4dk_z = 2\pi\quad.%
\end{equation}
Note that $\rho$ is a function of both $\omega$ and $k_z$. The solution of this equation is a dispersion relation $k_z(\omega)$, which can be converted to the more intuitive relation $k_x(\omega)$ by use of $k_x^2+k_z^2=\epseff\mueff (\omega/c)^2$. The resulting function $\omega(k_x)$ is indicated by the blue line in Fig.~\ref{fig:dispersion}(a).

To check the validity of our simple model, we performed detailed numerical calculations using \emph{CST Microwave Studio}. Moreover, to obtain an accurate reference to the experiments described later in this letter, we determined the dimensions of the fabricated metamaterial structure from a microscope image.
The numerically calculated dispersion diagram in Fig.~\ref{fig:dispersion}(a) (red crosses) shows two branches in the frequency range from 0 to \unit[2]{THz} in very good agreement with the results derived from the waveguide model. It is notable that the waveguide model enables us to fully describe the subwavelength-thick meta-surface as an effective medium with effective material parameters.

Besides the dispersion relation, we determined the magnetic field distribution of the guided modes from the numerical data. The results in Figs.~\ref{fig:dispersion}(b) and (c) reveal that the magnetic field distributions of the two mode branches in the dispersion relation significantly differ near the plane of the split rings. At a given wave vector $k_x$, the magnetic fields inside the rings and in the space between them oscillate in phase for the lower frequency mode, whereas they oscillate out of phase in the higher frequency mode. In the latter case, this leads to a stronger confinement of the fields in the immediate vicinity of the split rings since the fields cancel out with increasing distance from the plane.

\begin{figure}%
\centering%
\includegraphics{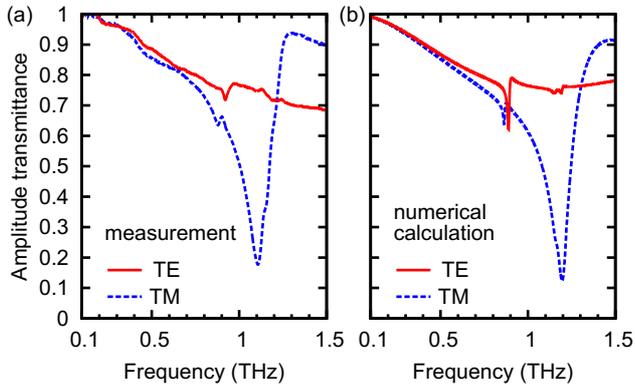}%
\caption{(a) Measured and (b) calculated spectra of terahertz transmission through the sample.}%
\label{fig:transmission}%
\end{figure}

To support the results by experimental data, we measured the transmission through the meta-surface using a terahertz time-domain spectroscopy setup under normal incidence. Figure \ref{fig:transmission} displays the measured spectra together with the numerically calculated results. The pronounced transmittance minimum at approx.\ \unit[1.1]{THz} for transverse magnetic (TM) polarization (magnetic field vector in $y$-direction) results from the electric field coupling to the fundamental resonance of the split rings. For TE polarization (electric field vector in $y$-direction), one sharp resonance feature appears at approx.\ \unit[0.9]{THz} as well as two weak dips at \unit[1.15]{THz} and \unit[1.2]{THz}. The frequencies of the transmission minima are directly related to the corresponding eigenfrequencies of the excited surface waves with wave vectors equal to integer multiples of the reciprocal grating vector, $k_x=m\times 2\pi/(6a)$. The stronger minimum at \unit[0.9]{THz} corresponds to the excitation of a lower mode surface wave with $m=1$ while the two weaker minima at \unit[1.15]{THz} and \unit[1.2]{THz} correspond to a lower mode surface wave with $m=2$ and a higher mode surface wave with $m=1$, respectively (see Fig.~\ref{fig:dispersion}).

In conclusion, we have experimentally demonstrated the excitation of resonant terahertz surface modes on thin magnetic metamaterials by transmission measurements. The observed transmission minima could be related to the excitation of surface waves by direct comparison to 3-D full-wave numerical calculations. In addition, we developed a simple and intuitive waveguide model to describe subwavelength-thick metamaterial films as an effective medium. From this model, we could deduce the dispersion relation of the surface waves in excellent agreement with the numerical and experimental results. The physical insight provided by this work is important with respect to the development of design tools for tailored surface waves on meta-surfaces.

\end{document}